\documentclass[prl,showpacs,twocolumn,amsfonts,amssymb,amsmath]{revtex4}
\usepackage{amsmath}
\usepackage{graphicx}
\usepackage{subfigure}
\usepackage{color}

\def \be {\begin{equation}}
\def \ee {\end{equation}}

\begin{document}

\title{Gutenberg-Richter Scaling - A New Paradigm}

\author{C. A. Serino}
\affiliation{Department of Physics, Boston
  University, Boston, MA, USA~ 02215}
\email{cserino@physics.bu.edu}
\author{K. F. Tiampo}
\affiliation{Department of Earth Sciences, University of Western Ontario, London, Ontario, Canada~ N6A~3K7}
\email{ktiampo@uwo.ca}
\author{W. Klein} 
\affiliation{Department of Physics and Center for Computational Science, Boston
  University, Boston, MA, USA~ 02215}
\email{klein@bu.edu}

\date{\today}

\begin{abstract}
We introduce a new model for an earthquake fault system that is composed of non-interacting simple lattice models with different levels of damage denoted by $q$. The undamaged lattice models ($q=0$) have Gutenberg-Richter scaling with a cumulative exponent $\beta=1/2$, whereas the damaged models do not have well defined scaling. However, if we consider the ``fault system'' consisting of all models, damaged and undamaged, we get excellent scaling with the exponent depending on the relative frequency with which faults with a particular amount of damage occur in the fault system.  This paradigm combines the idea that Gutenberg-Richter scaling is associated with an underlying critical point with the notion that the structure of a fault system also affects the statistical distribution of earthquakes. In addition, it provides a framework in which the variation, from one tectonic region to another, of the scaling exponent, or $b$-value, can be understood.
\end{abstract}
\maketitle

It has been known for over half a century that the frequency-magnitude distribution of seismic events occurring on a fault system obeys the Gutenberg-Richter (GR) scaling relation~\cite{GR,GutenbergBook} which states that magnitudes are distributed according to
\be
N_\mu\sim10^{-b\mu}\,,
\label{eq.GRorig}
\ee
where $\mu$ is the local or Richter magnitude~\cite{RichterBSSA35}. Today,  the Richter magnitude has been superseded by the more physically significant moment magnitude, $\mu_\text{w}$,~\cite{HanksJGR79} which is related to the average slip, $\overline{\Delta u}$, over the rupture surface by
\be
\mu_\text{w}=\frac{2}{3}\log_{10}M-10.7\,,\hspace{1mm}\text{where}\hspace{2mm}M=GA\overline{\Delta u}\,,
\label{eq.momentmagnitude}
\ee
$A$ is the area of and $G$ is the shear modulus of the rupture surface, and $M$ is the so-called seismic moment. The exponential distribution in the magnitude then becomes a power-law distribution in the moment, transforming Eq.~\ref{eq.GRorig} into
\be
N_M\sim M^{-\beta}\,,\hspace{1mm}\text{with}\hspace{2mm}\beta\equiv\frac{2b}{3}\,.
\label{eq.GRmom}
\ee
The distribution $N_M$ is the total number of events with moment {\it greater than or equal to} $M$ ~\cite{GR,GutenbergBook}. 
The inverse characteristic magnitude, $b$, in Eq.~\ref{eq.GRorig} is distributed in the narrow range $0.75\leq b \leq1.2$~\cite{runetal, gulia}. This range of $b$ values translates into a range $0.5\leq\beta\leq0.8$ for the exponent of the moment distribution.

The GR distribution in Eqs.~\ref{eq.GRorig}~and~\ref{eq.GRmom} has been the subject of considerable investigation by seismologists as well as physicists and geo-physicists who have associated the scaling with an underlying critical point. 
This apparent critical phenomena has led to the creation of several simple models that exhibit a GR-like distribution. The first, proposed by Burridge and Knopoff~\cite{BK}, consists of blocks connected by linear springs resting on a rough surface and is evolved according to Newton's Laws. Using the Mohr-Coulomb friction law and ignoring inertial effects, Rundle and Jackson~\cite{RJB} introduced a cellular automaton version of the Burridge-Knopoff model which was reintroduced in lattice form by Olami {\it et al.} (OFC)~\cite{OFC}. In addition to the association with critical points, Bak {\it et al.}~\cite{sandpile} and Feder and Feder~\cite{federfeder} proposed that GR scaling is a form of self organized criticality. Klein {\it et al.}~\cite{nucleationEQ} proposed that the scaling is associated with fluctuations about a spinodal critical point. All of these models share at least one shortcoming: they are best understood as describing individual faults rather than fault systems while it is the latter which are observed to scale according to Eqs.~\ref{eq.GRorig}~and~\ref{eq.GRmom}. Seismic events on single faults may or may not be described by the GR distribution, depending on the fault, as we will show later.

In this letter, we propose a different approach that models a {\it fault system} as a collection of non-interacting OFC models or ``faults'' with varying degrees of damage. The frequency of events on {\it individual} faults does not, in general, scale as a power-law in the event's moment, however, when a collection of model faults with varying degrees of damage
are assembled into a ``fault system", the aggregate distribution of moments is well described by the GR scaling relation.
The OFC model consists of a regular lattice of linear size $L$ with open boundary conditions and a stress failure threshold, $\sigma^{F}$, and a residual stress, $\sigma^{R}$, at each vertex (site). Initially, the sites are assigned a random stress, $\sigma^R \leq \sigma_j < \sigma^F$, where the index $j$ labels the position of the site on the lattice. The model is evolved via a plate update in which stress is simultaneously added to each site until the stress on the site nearest its failure threshold reaches $\sigma^{F}$. This loading method is the so-called zero velocity limit~\cite{OFC}. When a site fails its stress is reduced to $\sigma^R$ and  $(1 - \alpha)(\sigma_{j} - \sigma^{R}) / ar^2$ is distributed to each of its $ar^2$ neighbors  where $a$ is a dimensionless number depending on the shape of the interaction, $r$ is the range of the interaction, and $0\leq\alpha<1$ is the dissipation parameter. Failure of a site may cause neighboring sites to fail, resulting in an earthquake or avalanche. We continue to search the lattice until $\sigma_{j}<\sigma^{F}$~$\forall$~$j$, at which point the lattice is reloaded as described above.

We add a small amount of noise to the model by resetting each site to $\sigma^R\pm\eta$, where $\eta$ is drawn from a flat distribution with mean $\sigma^R$. For this paper, we take $\sigma^F=2$, $\sigma^R=1\pm0.025$, $L=512$, and stress from a failed site is redistributed within a circle of radius $r=20$. We choose $r=20$ to mimic the long-range stress transfer in fault systems~\cite{Rybicki86, Steketee58}. We simulate the system for $4\times10^6$ plate updates to remove any transient effects before recording the size (number of failed sites) of each of the next $10^7$ events. 

For nearest neighbor stress transfer ($r=1$), there is no apparent GR scaling~\cite{eqbook, grassberger,socolar}, however, if we take $r\gg 1$, the system becomes near meanfield~\cite{eqbook} and there appears to be GR or cluster scaling associated with a spinodal critical point~\cite{nucleationEQ, eqbook}. In this limit, the number of events, $n_s$, of size $s \in [s\,,s+\text{d}s]$ scales as
\be
\label{eq.spinscale}
n_{s} = n_0\:\frac{\exp \left[-\Delta h\: s\right]}{s^{\tau}}\,,
\ee     
with $\tau = 3/2$. Here, $n_0$ is a measure of the seismic activity of the fault and $\Delta h$ is a measure of the distance from the spinodal~\cite{sof, eqbook}. To make contact with observations on real faults, we note that for $r\gg1$, the size, $s$ in 
Eq.~\ref{eq.spinscale}, is proportional to the moment~\cite{eqbook}. To obtain the cumulative distribution, $N_{M}$, we simply integrate
\be
N_M\propto N_s=\int_{s}^{\infty}\text{d}\xi\,n_\xi\,.
\label{eq.d2c}
\ee

\begin{figure}[!htbp]
\begin{center}
\includegraphics[width=0.48\textwidth]{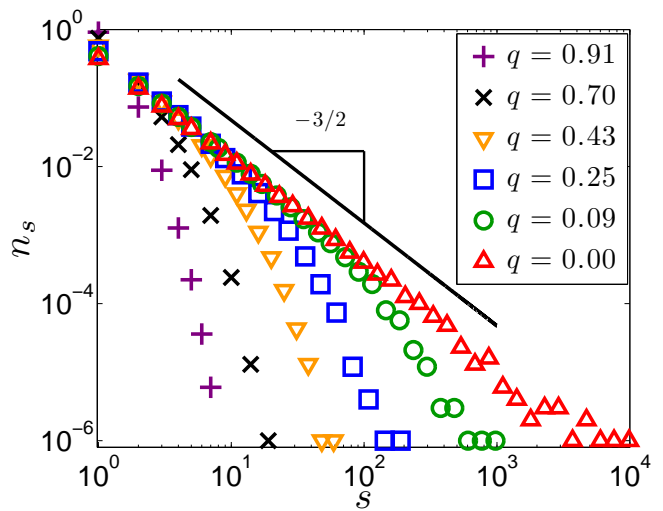}
\caption{(color online) The frequency-size distribution for a model fault for various values of $q$. For $q=0$ (i.e. no damage) we observe scaling up to the linear system size. As $q$ is increased, the distribution is no longer described by a power-law, rather, large events are exponentially suppressed according to Eq.~\ref{eq.spinscale}. The solid black line has slope $-3/2$, the exponent predicted for the scaling on undamaged faults. To see the distributions clearly, we do not plot $n_s$ for every value of $s$. Instead, we plot the distribution at evenly spaced values of $s$ in log-space. All distributions contain the same number of events.}
\label{fig.if}
\end{center}
\end{figure}

We now modify the model by randomly eliminating a fraction of sites, $q$~\cite{serinoPRE10}. These eliminated or dead sites do not hold stress, hence, when stress is transferred to a dead site from a failing neighbor, it is dissipated. In real faults, this can be caused by different levels of fracture or gouge, which are known to vary over a fault system~\cite{sammis}. We do this for many lattices with $0\leq q < 1$. As can be seen from Fig~\ref{fig.if}, the larger the value of $q$ the less the distribution resembles a simple power-law.
This is consistent with the idea that the OFC model in the zero velocity limit is only critical with a zero dissipation constant~\cite{henrik}. Nevertheless, the data is still well described by Eq.~\ref{eq.spinscale}, but with $q$-dependent $\Delta h$ and $n_0$. 

\begin{figure}[!htbp]
\begin{center}
\includegraphics[width=0.48\textwidth]{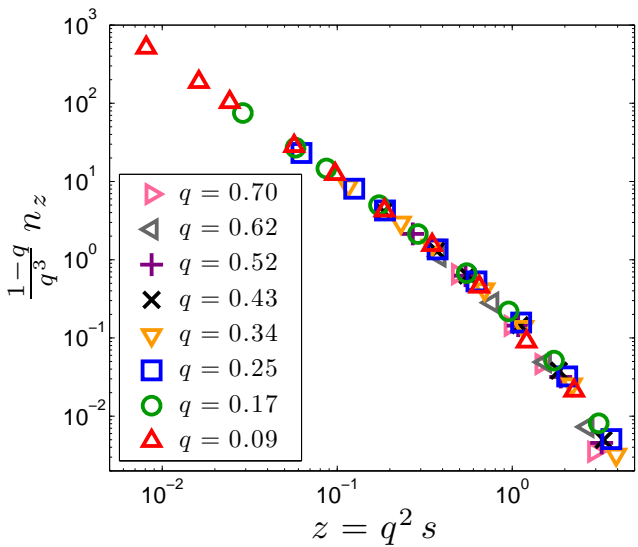}
\caption{(color online) The frequency-size distribution plotted in terms of the scaling variables. We explicitly account for the $q$-dependence in $n_s$ by plotting $(1-q)/q^3\,n_z$ versus $z=q^2s$. Having removed the $q$-dependence in these distributions, the data collapses to a single curve for all values of $q$. To see the collapse clearly, we plot the distribution at evenly spaced values of $s$ in log-space. All distributions contain the same number of events.}
\label{fig.dc}
\end{center}
\end{figure}

To determine the dependence of  $\Delta h$ and $n_0$ on $q$, we consider the theoretical description of the model as derived by Klein {\it et al.}~\cite{nucleationEQ,eqbook} from which one can argue for the dependence
\be
\label{eq.damage}
n_{s} =  \frac{N}{1-q}\frac{\exp \left[-q^{2}\:  s\right]}{s^{\tau}}\,,
\ee
where $\tau = 3/2$, as above, and $N$ is a $q$-independent constant related to the average seismic activity on the fault. We check this $q$-dependence in two ways. First, for various values of $q$, we fit $n_s$, using a least-squares, non-linear, weighted fit, for $\Delta h$ and $n_0$ to Eq.~\ref{eq.damage}. Additionally, we introduce a scaling variable $z\equiv q^2s$ such that plots of $n_z\,(1-q)/q^3$ should collapse to a single curve for all values of $q$. The data collapse in Fig.~\ref{fig.dc} suggests we have the correct scaling form for $n_s$.

\begin{figure}[!htbp]
\begin{center}
\includegraphics[width=0.48\textwidth]{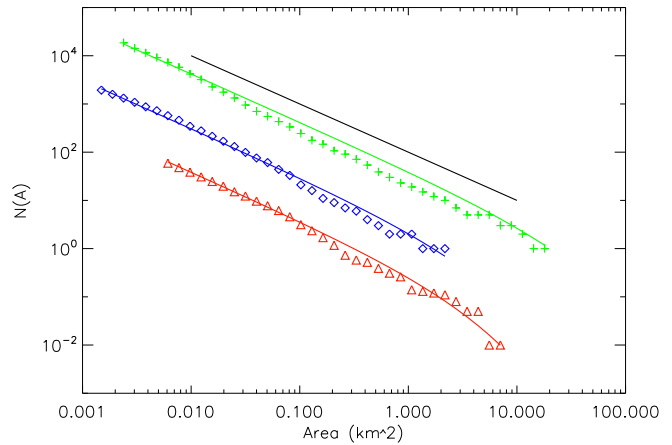}
\caption{(color online) The number of events with magnitude greater than or equal to the minimum magnitude of completeness~\cite{WiemerWyss00} versus the area of each event~\cite{WellsCoppersmith94} occurring from 1980~-~2008 within a 20 km swatch on either side of the San Jacinto fault (green crosses), Ft. Tejon segment of the San Andreas Fault (blue diamonds), and the Creeping section of the San Andreas Fault (red triangles). To see the data clearly, the various faults have been offset by factors of ten. The fits (solid colored lines) are to Eq.~\ref{eq.GRarea} and the solid black line has unit slope.}
\label{fig.socal}
\end{center}
\end{figure}

If this mechanism is responsible for the frequency-magnitude distribution on {\it single faults}, then we expect the data obtained from {\it real} faults to vary from fault to fault but, this variation should be captured by a single parameter in an otherwise universal distribution, as in Eq.~\ref{eq.spinscale}. Using scaling arguments, one can rewrite Eq.~\ref{eq.spinscale} in terms of the slip area, $A$, introduced in Eq.~\ref{eq.momentmagnitude}, as 
\be
\label{eq.GRarea}
N_{A}\propto \frac{\exp\left[-\Delta h A^{2/3}\right]}{A}\,.
\ee
In Fig.~\ref{fig.socal}, we plot the cumulative number of events with area $A$ on three faults from the southern California system.  
As can be seen from the figure, the data is consistent with Eq.~\ref{eq.GRarea}.

\begin{figure}[!htbp]
\begin{center}
\includegraphics[width=0.48\textwidth]{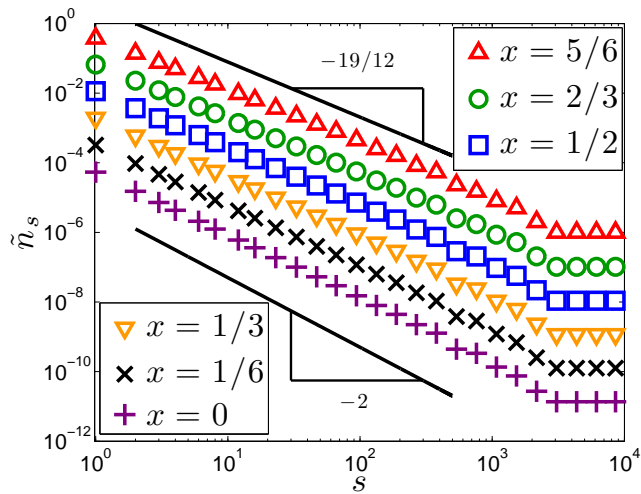}
\caption{(color online) The cumulative frequency-size distribution for a model fault system for various damage distributions, $D_q\sim q^{-x}$. By varying the exponent $x$, the GR distribution scales with different exponents $\tilde{\tau}$. The solid black lines have slope $-19/12$ and $-2$, the exponents predicted by Eq.~\ref{eq.weightedscaling} for $x=5/6$ and $x=0$, respectively. To see the data clearly, we plot the distribution at evenly spaced values of $s$ is log-space and we offset different values of $x$ by scaling the data by factors of $10$. All distributions contain the same number of events.}

\label{fig.fs}
\end{center}
\end{figure}

\begin{figure}[!htbp]
\begin{center}
\includegraphics[width=0.48\textwidth]{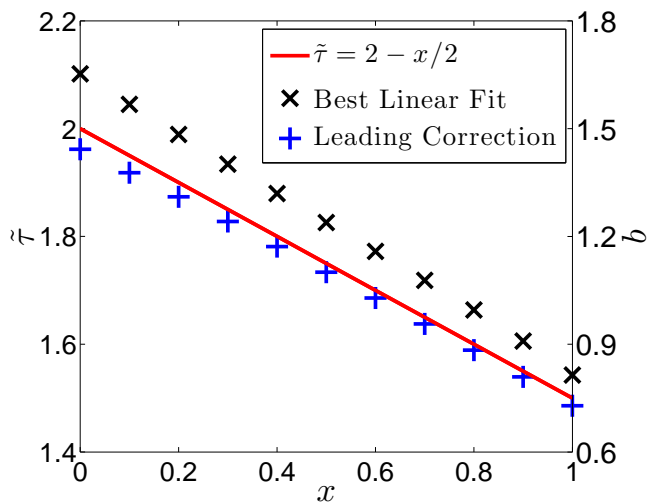}
\caption{(color online) The scaling exponents, $\tilde{\tau}$ and $b$ (c.f Eqs.~\ref{eq.GRorig}~and~\ref{eq.sum}, respectively), versus the damage distribution exponent, $x$. The black exes represent the value of $\tilde{\tau}$ as determined by fitting the data to a single power-law. The blue crosses are the value of $\tilde{\tau}$ determined by fitting the data to a power-law plus the leading asymptotic correction to the scaling form. The solid red line is the value of $\tilde{\tau}$ predicted by Eq.~\ref{eq.weightedscaling}.}
\label{fig.exp}
\end{center}
\end{figure}

To obtain the frequency-size distribution for a {\it fault system}, we must integrate Eq.~\ref{eq.damage} over all values of $q$, that is,
\be
\label{eq.sum}
\tilde{n}_s= \int_0^1\text{d}q\,n_s\sim\int_{0}^{1}\text{d}q\,\frac{D_q}{1-q} \frac{\exp \left[-q^2\: s\right]}{s^\tau}\sim\frac{1}{s^{\tilde{\tau}}}\,,
\ee
where the weighting factor, $D_q$, is the fraction of faults with damage $q\in[q\,,\,q+\text{d}q]$. For equal weighting ($D_q = 1$) we find $\tilde{n}_s\sim 1/s^{2}$ which implies a cumulative exponent $\beta = \tilde{\tau} - 1= 1$ (c.f. Eq.~\ref{eq.d2c}).
 In Fig.~\ref{fig.fs}, we plot (purple crosses) the number of events of a given size versus $s$. 
 As noted above, $\beta$ is observed to vary between $0.5$ and $0.8$ in real fault systems~\cite{runetal, gulia}, however, in real fault systems we do not know $D_q$. If we associate the damage with micro-cracks, which are know to have a fractal spatial distribution~\cite{sammis}, it is reasonable to assume a scale-free or power-law distribution for $D_q\sim q^{-x}$, where the exponent varies depending on the fault system. Therefore, the frequency-size distribution in Eq.~\ref{eq.sum} can be written in terms of the incomplete gamma function. If we perform an asymptotic expansion in the limit $s\gg1$, we find
\be
\tilde{n}_s\sim\frac{1}{s^{2-x/2}}\left(1+\frac{\Gamma[(2-x)/2]}{\Gamma[(1-x)/2]}\frac{1}{s^{1/2}}+\mathcal{O}\left(s^{-1}\right)\right)\,,
\label{eq.weightedscaling}
\ee
where $\Gamma(z)$ is the gamma function. To leading order, we find $\tilde{n}_s\sim s^{-\tilde{\tau}}$ with $\tilde{\tau}=2-x/2$, and so if we wish to obtain a $b$-value of unity, we simply solve
\be
b=\frac{3(2-x)}{4}\,,
\label{eq.bofx}
\ee
with $b=1$, so $x=2/3$. Furthermore, it is clear from Eq.~\ref{eq.bofx} that a range of $b$ values can be obtained by varying the exponent in $D_q$~\footnote{Note that $x$ is restricted to $x<1$ for the integral to converge, however, physically observed $b$-values correspond to the range $0.4 < x < 1$.} as shown in Fig~\ref{fig.fs}. 
In Fig~\ref{fig.exp}, we plot the the scaling exponents $\tilde{\tau}$ and $b$ versus the exponent, $x$, of the damage distribution, $D_q$. 
The solid red line is the relationship predicted by Eq.~\ref{eq.weightedscaling} while the black exes represents the value of $\tilde{\tau}$ determined by a linear fit of the data to the form $N/s^{\tilde{\tau}}$ for $N$ and $\tilde{\tau}$. To get better agreement between this fit value and our predicted value, we perform a two parameter non-linear fit to the form $N/s^{\tilde{\tau}}(1+\gamma/s^{1/2})$ where $\gamma = \Gamma[(2-x)/2] / \Gamma[(1-x)/2]$ is not a fit parameter. The value of $\tilde{\tau}$ as determined from this fit is shown in blue crosses. 

In summary, we have introduced a new paradigm for understanding GR scaling 
which links scaling associated with a (spinodal) critical point on a single fault with GR scaling on a fault system. The GR scaling on a fault system is affected by both the degree of damage on each fault ($q$), and the distribution of damage over the fault system ($D_q$). This paradigm offers an explanation for the observation of GR scaling on fault systems comprised of individual faults: none of which produce the same event magnitude statistics as the aggregate system. It also provides a framework in which one can understand the observed variation of the GR exponent, or $b$-value, from one region to another. In addition, since the various faults that comprise the fault system are non-interacting by construction, our model provides a mechanism in which scaling can arise on fault system length scales that are ten to one-hundred times larger than any length scale associated with the critical phenomena occurring on a single fault.

Two of the authors (C.A.S and W.K.) wish to thank the DOE for support through grant DE-FG02-95ER14498 and one of the authors (K.F.T) would like to thank the NSERC and Aon Benfield/ICLR Industrial Research Chair in Earthquake Hazard Assessment as well as an NSERC Discovery Grant.
\bibliography{GRbvalues}

 \end{document}